\documentclass[twocolumn,twocolappendix]{aastex631}
\usepackage{xcolor}
\usepackage{amsmath}
\usepackage{stmaryrd}
\usepackage{amssymb}
\usepackage{wasysym}

\newcommand{\RNum}[1]{\uppercase\expandafter{\romannumeral #1\relax}}
\usepackage{subfigure}

\usepackage{appendix}

\begin{document}
\title{Detection of optical quasi-periodic oscillation in the blazar 3C 454.3}

\author[0009-0007-3214-602X]{Karan Dogra}\email{karandogra987@gmail.com}
\affiliation{Aryabhatta Research Institute of Observational Sciences(ARIES), Manora Peak, Nainital 263001, India}
\affiliation{Department of Applied Physics, Mahatma Jyotiba Phule Rohilkhand University, Bareilly 243006, India}

\author[0000-0002-9331-4388]{Alok C.\ Gupta}\email{acgupta30@gmail.com}
\affiliation{Aryabhatta Research Institute of Observational Sciences(ARIES), Manora Peak, Nainital 263001, India}
\affiliation{Xinjiang Astronomical Observatory, Chinese Academy of Sciences, 150 Science-1 Street, Urumqi 830011, China}
\affiliation{Abastumani Observatory, Mt. Kanobili, 0301 Abastumani, Georgia}

\author[0000-0003-1784-2784]{C.~M.~Raiteri}
\affiliation{INAF, Osservatorio Astrofisico di Torino, via Osservatorio 20, I-10025 Pino Torinese, Italy}

\author[0000-0003-1743-6946]{M.~Villata}
\affiliation{INAF, Osservatorio Astrofisico di Torino, via Osservatorio 20, I-10025 Pino Torinese, Italy}

\author[0000-0002-1029-3746]{Paul J.~Wiita}
\affiliation{Department of Physics, The College of New Jersey, PO Box 7718, Ewing, NJ 08628, USA}

\author[0000-0001-8580-8874]{Mauri J.~Valtonen}
\affiliation{FINCA, University of Turku, FI-20014 Turku, Finland}
\affiliation{Tuorla Observatory, Department of Physics and Astronomy, University of Turku, FI-20014 Turku, Finland}

\author[0000-0002-0319-5873]{S. O. Kurtanidze}
\affiliation{Abastumani Observatory, Mt. Kanobili, 0301 Abastumani, Georgia}

\author[0000-0001-6158-1708]{S. G. Jorstad}
\affiliation{Institute for Astrophysical Research, Boston University, 725 Commonwealth Avenue, Boston, MA 02215, USA}
\affiliation{Astronomical Institute, Saint Petersburg State University, 7/9 Universitetskaya nab., St. Petersburg, 199034, Russia}

\author[0000-0002-0766-864X]{R. Bachev}
\affiliation{Institute of Astronomy and National Astronomical Observatory, Bulgarian Academy of Sciences, 72 Tsarigradsko shosse Blvd., 1784 Sofia, Bulgaria}

\author[0000-0002-6710-6868]{G. Damljanovic}
\affiliation{Astronomical Observatory, Volgina 7, 11060 Belgrade, Serbia}

\author[0009-0002-5220-2993]{C. Lorey}
\affiliation{Hans-Haffner-Sternwarte, Naturwissenschaftliches Labor f\"{u}r Sch\"{u}ler am FKG, Friedrich-Koenig-Gymnasium, D-97082 W$\ddot{u}$rzburg, Germany}

\author[0000-0003-4147-3851]{S. S. Savchenko}
\affiliation{Astronomical Institute, Saint Petersburg State University, 7/9 Universitetskaya nab., St. Petersburg, 199034, Russia}
\affiliation{Special Astrophysical Observatory, Russian Academy of Sciences, 369167, Nizhnii Arkhyz, Russia}
\affiliation{Pulkovo Observatory, St.Petersburg, 196140, Russia}

\author[0009-0008-5761-3701]{O. Vince}
\affiliation{Astronomical Observatory, Volgina 7, 11060 Belgrade, Serbia}

\author{M. Abdelkareem}
\affiliation{National Research Institute of Astronomy and Geophysics (NRIAG), 11421 Helwan, Cairo, Egypt}

\author{F. J. Aceituno}
\affiliation{Instituto de Astrof\'{i}sica de Andaluc\'{i}a, IAA-CSIC, Glorieta de la Astronom\'{i}a s/n, E-18008 Granada, Spain}

\author[0000-0002-0433-9656]{J. A. Acosta-Pulido}
\affiliation{Instituto de Astrof\'{i}sica de Canarias (IAC), E-38200 La Laguna, Tenerife, Spain}
\affiliation{Universidad de La Laguna, Departamento de Astrof\'{i}sica, E-38206 La Laguna, Tenerife, Spain}

\author[0000-0002-3777-6182]{I. Agudo}
\affiliation{Instituto de Astrof\'{i}sica de Andaluc\'{i}a, IAA-CSIC, Glorieta de la Astronom\'{i}a s/n, E-18008 Granada, Spain}

\author[0000-0001-5125-6397]{G. Andreuzzi}
\affiliation{INAF, TNG Fundaci{\'o}n Galileo Galilei, La Palma, E-38712, Spain}

\author{S. A. Ata}
\affiliation{National Research Institute of Astronomy and Geophysics (NRIAG), 11421 Helwan, Cairo, Egypt}

\author{G. V. Baida}
\affiliation{Crimean Astrophysical Observatory of the Russian Academy of Sciences, P/O Nauchny 298409, Russia}

\author{L. Barbieri}
\affiliation{Orciatico Astronomical Observatory, Orciatico (Pisa), Italy}

\author[0000-0003-0611-5784]{D. A. Blinov}
\affiliation{Institute of Astrophysics, Foundation for Research and Technology - Hellas, Voutes, 70013 Heraklion, Greece}
\affiliation{Department of Physics, University of Crete, 71003, Heraklion, Greece}

\author{G. Bonnoli}
\affiliation{Instituto de Astrof\'{i}sica de Andaluc\'{i}a, IAA-CSIC, Glorieta de la Astronom\'{i}a s/n, E-18008 Granada, Spain}
\affiliation{INAF Osservatorio Astronomico di Brera, Via E. Bianchi 46, 23807 Merate (LC), Italy}

\author[0000-0002-7262-6710]{G. A. Borman}
\affiliation{Crimean Astrophysical Observatory of the Russian Academy of Sciences, P/O Nauchny 298409, Russia}

\author[0000-0001-5843-5515]{M. I. Carnerero}
\affiliation{INAF, Osservatorio Astrofisico di Torino, via Osservatorio 20, I-10025 Pino Torinese, Italy}

\author[0000-0001-5252-1068]{D. Carosati}
\affiliation{EPT Observatories, Tijarafe, La Palma, E-38780, Spain}
\affiliation{INAF, TNG Fundaci{\'o}n Galileo Galilei, La Palma, E-38712, Spain}

\author[0000-0003-2036-8999]{V. Casanova}
\affiliation{Instituto de Astrof\'{i}sica de Andaluc\'{i}a, IAA-CSIC, Glorieta de la Astronom\'{i}a s/n, E-18008 Granada, Spain}

\author[0000-0003-0262-272X]{W. P. Chen}
\affiliation{Institute of Astronomy, National Central University, Taoyuan 32001, Taiwan}

\author[0000-0003-0721-5509]{Lang Cui}
\affiliation{Xinjiang Astronomical Observatory, Chinese Academy of Sciences, 150 Science-1 Street, Urumqi 830011, China}

\author[0000-0003-3337-4861]{P. U. Devanand}
\affiliation{Aryabhatta Research Institute of Observational Sciences(ARIES), Manora Peak, Nainital 263001, India}
\affiliation{Department of Applied Physics, Mahatma Jyotiba Phule Rohilkhand University, Bareilly 243006, India}

\author[0000-0002-9751-8089]{E. G. Elhosseiny}
\affiliation{National Research Institute of Astronomy and Geophysics (NRIAG), 11421 Helwan, Cairo, Egypt}

\author[0000-0001-6796-3205]{D. Elsaesser}
\affiliation{Hans-Haffner-Sternwarte, Naturwissenschaftliches Labor f\"{u}r Sch\"{u}ler am FKG, Friedrich-Koenig-Gymnasium, D-97082 W$\ddot{u}$rzburg, Germany}
\affiliation{Astroteilchenphysik, TU Dortmund, Otto-Hahn-Str. 4A, D-44227 Dortmund, Germany}

\author[0000-0002-4131-655X]{J. Escudero}
\affiliation{Center for Astrophysics | Harvard \& Smithsonian, Cambridge, MA 02138, USA}
\affiliation{Instituto de Astrof\'{i}sica de Andaluc\'{i}a, IAA-CSIC, Glorieta de la Astronom\'{i}a s/n, E-18008 Granada, Spain}

\author[0000-0002-5929-0968]{J. H. Fan}
\affiliation{Center for Astrophysics, Guangzhou University, Guangzhou 510006, People's Republic of China}
\affiliation{Astronomy Science and Technology Research Laboratory of the Department of Education of Guangdong Province, Guangzhou 510006, People's Republic of China}
\affiliation{Greater Bay Brand Center of the National Astronomical Data Center, Guangzhou 510006, People's Republic of China}

\author{M. Feige}
\affiliation{Hans-Haffner-Sternwarte, Naturwissenschaftliches Labor f\"{u}r Sch\"{u}ler am FKG, Friedrich-Koenig-Gymnasium, D-97082 W$\ddot{u}$rzburg, Germany}

\author[0000-0002-8855-3923]{K. Gazeas}
\affiliation{Section of Astrophysics, Astronomy and Mechanics, Department of Physics, National and Kapodistrian University of Athens, GR-15784 Zografos, Athens, Greece}

\author{T. S. Grishina}
\affiliation{Astronomical Institute, Saint Petersburg State University, 7/9 Universitetskaya nab., St. Petersburg, 199034, Russia}

\author[0000-0002-4455-6946]{Minfeng Gu}
\affiliation{Shanghai Astronomical Observatory, Chinese Academy of Sciences, 80 Nandan Road, Shanghai 200030, China}

\author{V. A. Hagen-Thorn}
\affiliation{Astronomical Institute, Saint Petersburg State University, 7/9 Universitetskaya nab., St. Petersburg, 199034, Russia}

\author{F. Hemrich}
\affiliation{Hans-Haffner-Sternwarte, Naturwissenschaftliches Labor f\"{u}r Sch\"{u}ler am FKG, Friedrich-Koenig-Gymnasium, D-97082 W$\ddot{u}$rzburg, Germany}

\author{H. Y. Hsiao}
\affiliation{Institute of Astronomy, National Central University, Taoyuan 32001, Taiwan}

\author{M. Ismail}
\affiliation{National Research Institute of Astronomy and Geophysics (NRIAG), 11421 Helwan, Cairo, Egypt}

\author[0009-0005-7297-8985]{R. Z. Ivanidze}
\affiliation{Abastumani Observatory, Mt. Kanobili, 0301 Abastumani, Georgia}

\author[0000-0003-4298-3247]{M. D. Jovanovic}
\affiliation{Astronomical Observatory, Volgina 7, 11060 Belgrade, Serbia}

\author{T. M. Kamel}
\affiliation{National Research Institute of Astronomy and Geophysics (NRIAG), 11421 Helwan, Cairo, Egypt}

\author[0000-0002-5684-2114]{G. N. Kimeridze}
\affiliation{Abastumani Observatory, Mt. Kanobili, 0301 Abastumani, Georgia}

\author[0000-0001-8716-9412]{Shubham Kishore}
\affiliation{Aryabhatta Research Institute of Observational Sciences(ARIES), Manora Peak, Nainital 263001, India}
\affiliation{Indian Institute of Astrophysics (IIA), 2nd Block, Koramangala, Bangalore 560034, India}

\author{E. N. Kopatskaya}
\affiliation{Astronomical Institute, Saint Petersburg State University, 7/9 Universitetskaya nab., St. Petersburg, 199034, Russia}

\author{D. Kuberek}
\affiliation{Hans-Haffner-Sternwarte, Naturwissenschaftliches Labor f\"{u}r Sch\"{u}ler am FKG, Friedrich-Koenig-Gymnasium, D-97082 W$\ddot{u}$rzburg, Germany}

\author[0000-0001-5385-0576]{O. M. Kurtanidze}
\affiliation{Abastumani Observatory, Mt. Kanobili, 0301 Abastumani, Georgia}
\affiliation{Engelhardt Astronomical Observatory, Kazan Federal University, Tatarstan, Russia}
\affiliation{Landessternwarte, Zentrum für Astronomie der Universität Heidelberg, Königstuhl 12, 69117 Heidelberg, Germany}

\author{A. Kurtenkov}
\affiliation{Institute of Astronomy and National Astronomical Observatory, Bulgarian Academy of Sciences, 72 Tsarigradsko shosse Blvd., 1784 Sofia, Bulgaria}

\author{V. M. Larionov}
\affiliation{Astronomical Institute, Saint Petersburg State University, 7/9 Universitetskaya nab., St. Petersburg, 199034, Russia}

\author[0000-0002-2471-6500]{Elena G. Larionova}
\affiliation{Astronomical Institute, Saint Petersburg State University, 7/9 Universitetskaya nab., St. Petersburg, 199034, Russia}

\author{L. V. Larionova}
\affiliation{Astronomical Institute, Saint Petersburg State University, 7/9 Universitetskaya nab., St. Petersburg, 199034, Russia}

\author{H. C. Lin}
\affiliation{Institute of Astronomy, National Central University, Taoyuan 32001, Taiwan}


\author[0000-0003-3779-6762]{A. Marchini}
\affiliation{University of Siena, Astronomical Observatory, Via Roma 56. 53100, Siena, Italy}

\author[0000-0002-3596-4307]{C. Marinelli}
\affiliation{Dip. di Scienze Fisiche, della Terra e dell’Ambiente, Universit\'{a} di Siena, Via Roma 56, 53100, Siena, Italy}

\author[0000-0001-7396-3332]{A. P. Marscher}
\affiliation{Institute for Astrophysical Research, Boston University, 725 Commonwealth Avenue, Boston, MA 02215, USA}

\author{D. Morcuende}
\affiliation{Instituto de Astrof\'{i}sica de Andaluc\'{i}a, IAA-CSIC, Glorieta de la Astronom\'{i}a s/n, E-18008 Granada, Spain}

\author{D. A. Morozova}
\affiliation{Astronomical Institute, Saint Petersburg State University, 7/9 Universitetskaya nab., St. Petersburg, 199034, Russia}

\author{S. V. Nazarov}
\affiliation{Crimean Astrophysical Observatory of the Russian Academy of Sciences, P/O Nauchny 298409, Russia}

\author[0000-0003-0408-7177]{M. G. Nikolashvili}
\affiliation{Abastumani Observatory, Mt. Kanobili, 0301 Abastumani, Georgia}

\author{D. Reinhart}
\affiliation{Hans-Haffner-Sternwarte, Naturwissenschaftliches Labor f\"{u}r Sch\"{u}ler am FKG, Friedrich-Koenig-Gymnasium, D-97082 W$\ddot{u}$rzburg, Germany}

\author[0000-0002-4241-5875]{J. Otero-Santos}
\affiliation{Instituto de Astrof\'{i}sica de Andaluc\'{i}a, IAA-CSIC, Glorieta de la Astronom\'{i}a s/n, E-18008 Granada, Spain}

\author{A. Scherbantin}
\affiliation{Hans-Haffner-Sternwarte, Naturwissenschaftliches Labor f\"{u}r Sch\"{u}ler am FKG, Friedrich-Koenig-Gymnasium, D-97082 W$\ddot{u}$rzburg, Germany}

\author[0000-0002-1839-3936]{E. Semkov}
\affiliation{Institute of Astronomy and National Astronomical Observatory, Bulgarian Academy of Sciences, 72 Tsarigradsko shosse Blvd., 1784 Sofia, Bulgaria}

\author{E. V. Shishkina}
\affiliation{Astronomical Institute, Saint Petersburg State University, 7/9 Universitetskaya nab., St. Petersburg, 199034, Russia}

\author[0000-0002-6985-2143]{L. A. Sigua}
\affiliation{Abastumani Observatory, Mt. Kanobili, 0301 Abastumani, Georgia}

\author[0000-0001-7881-7748]{A. K. Singh}
\affiliation{Department of Applied Physics, Mahatma Jyotiba Phule Rohilkhand University, Bareilly 243006, India}

\author{A. Sota}
\affiliation{Instituto de Astrof\'{i}sica de Andaluc\'{i}a, IAA-CSIC, Glorieta de la Astronom\'{i}a s/n, E-18008 Granada, Spain}

\author{R. Steineke}
\affiliation{Hans-Haffner-Sternwarte, Naturwissenschaftliches Labor f\"{u}r Sch\"{u}ler am FKG, Friedrich-Koenig-Gymnasium, D-97082 W$\ddot{u}$rzburg, Germany}

\author[0000-0002-4105-7113]{M. Stojanovic}
\affiliation{Astronomical Observatory, Volgina 7, 11060 Belgrade, Serbia}

\author{A. Strigachev}
\affiliation{Institute of Astronomy and National Astronomical Observatory, Bulgarian Academy of Sciences, 72 Tsarigradsko shosse Blvd., 1784 Sofia, Bulgaria}

\author[0000-0003-1423-5516]{A. Takey}
\affiliation{National Research Institute of Astronomy and Geophysics (NRIAG), 11421 Helwan, Cairo, Egypt}

\author[0000-0002-8279-9236]{Amira A. Tawfeek}
\affiliation{National Research Institute of Astronomy and Geophysics (NRIAG), 11421 Helwan, Cairo, Egypt}

\author[0009-0006-3586-2489]{Tushar Tripathi}
\affiliation{Aryabhatta Research Institute of Observational Sciences(ARIES), Manora Peak, Nainital 263001, India}

\author{I. S. Troitskiy}
\affiliation{Astronomical Institute, Saint Petersburg State University, 7/9 Universitetskaya nab., St. Petersburg, 199034, Russia}

\author{Y. V. Troitskaya}
\affiliation{Astronomical Institute, Saint Petersburg State University, 7/9 Universitetskaya nab., St. Petersburg, 199034, Russia}

\author[0000-0002-3211-4219]{An-Li Tsai}
\affiliation{Institute of Astronomy, National Central University, Taoyuan 32001, Taiwan}
\affiliation{Department of Physics, National Sun Yat-sen University, Kaohsiung, 80424, Taiwan}

\author{A. A. Vasilyev}
\affiliation{Astronomical Institute, Saint Petersburg State University, 7/9 Universitetskaya nab., St. Petersburg, 199034, Russia}

\author[0009-0002-7669-7425]{K. Vrontaki}
\affiliation{Section of Astrophysics, Astronomy and Mechanics, Department of Physics, National and Kapodistrian University of Athens, GR-15784 Zografos, Athens, Greece}

\author[0000-0002-8366-3373]{Zhongli Zhang}
\affiliation{Shanghai Astronomical Observatory, Chinese Academy of Sciences, 80 Nandan Road, Shanghai 200030, China}
\affiliation{Key Laboratory of Radio Astronomy and Technology, Chinese Academy of Sciences, A20 Datun Road, Chaoyang District, Beijing 100101, China}

\author{A. V. Zhovtan}
\affiliation{Crimean Astrophysical Observatory of the Russian Academy of Sciences, P/O Nauchny 298409, Crimea}

\author{N. Zottmann}
\affiliation{Hans-Haffner-Sternwarte, Naturwissenschaftliches Labor f\"{u}r Sch\"{u}ler am FKG, Friedrich-Koenig-Gymnasium, D-97082 W$\ddot{u}$rzburg, Germany}

\author[0000-0002-4521-6281]{Wenwen Zuo}
\affiliation{Shanghai Astronomical Observatory, Chinese Academy of Sciences, 80 Nandan Road, Shanghai 200030, China}

\begin{abstract}
\noindent  We analyzed 19 years of $R$-band data of the blazar 3C 454.3 from the Whole Earth Blazar Telescope (WEBT) archive, along with new data from its members and from public archives such as those provided by the Small and Moderate Aperture Research Telescope System (SMARTS) and the Steward Observatory projects to search for quasi-periodic oscillations (QPOs). We detected a QPO of $\sim$ 433 days using Lomb-Scargle periodogram, which lasted from MJD 54980--58450 as detected by the weighted wavelet Z-transform technique, making it one of the most persistent QPOs ever detected in the optical regime. The phase dispersion minimization technique was also performed to further validate this QPO claim. We detected this signal at a global significance of $2.53\sigma$ across all methodologies.
To explain the observed QPO, we have considered both models focused on the accretion disk around the super-massive black hole (SMBH), and those based purely on jet emissions. Plausible jet-based models involve a shock moving down the jet in a helical magnetic field, whereas the SMBH models could involve Lense-Thirring effect-induced jet precession or dual jets in a binary SMBH system. We introduce a novel approach to distinguish genuine QPOs from spurious signals arising from annual seasonal gaps, a common limitation of ground-based observations.

\end{abstract}


\section{Introduction}\label{sec1}
The relativistic jets of blazars, a subgroup of active galactic nuclei (AGN), point roughly in the direction of our line of sight \citep{1995PASP..107..803U}. Blazars are divided into two subclasses: BL Lacertae objects (BL Lacs) and flat-spectrum radio quasars (FSRQs). In the composite near-infrared-to-UV  
spectra, strong emission lines are usually seen in FSRQs, while no significant emission or absorption lines are seen in BL Lacs. Accreting black holes (BHs) with masses ranging from $10^{6}-\rm{10}^{10} M_{\odot}$ are almost certainly present in AGNs, and they resemble scaled-up galactic X-ray producing BH binaries in many ways.

\begin{figure*}
    \centering
    \includegraphics[scale=0.8]{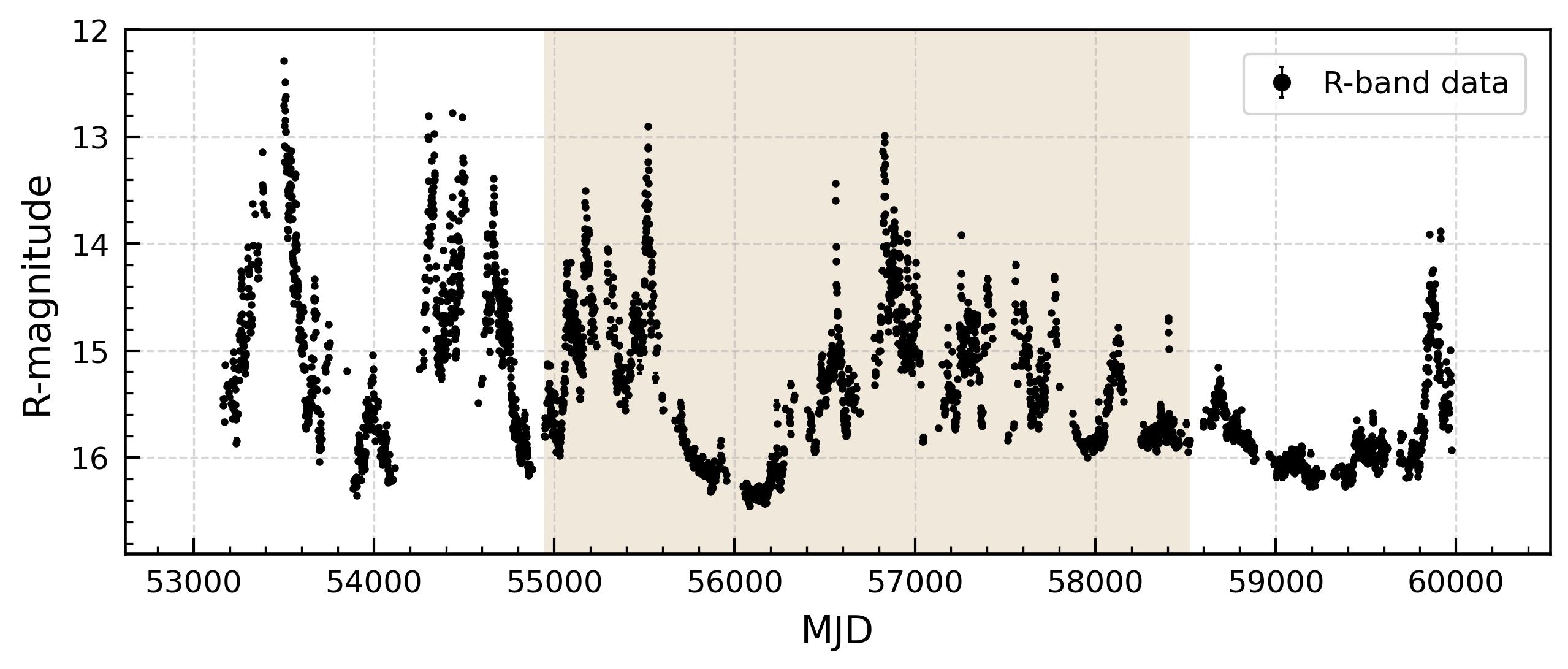}
    \caption{R-band light curve plot of the object 3C 454.3. The plot shows the full light curve from 2004 to 2023. The part of the light curve where the signal is most apparent is highlighted.}
    \label{R_band_zoomed}
\end{figure*}

In our and neighboring galaxies, quasi-periodic oscillations (QPOs) are fairly frequent in both BHs and neutron star binaries \citep[e.g.][]{2006ARA&A..44...49R}. However, the light curves (LCs) of AGNs are mostly non-periodic throughout the whole electromagnetic (EM) spectrum, exhibiting stochastic changes that can be ascribed to accretion disk (AD) or jet instabilities \citep[e.g.][and references therein]{2018A&A...616L...6G,2021MNRAS.501.5997T,2024MNRAS.527.9132T}. Nonetheless, sporadic QPO detections in various EM bands with varying timescales have been argued for in a number of blazars \citep[e.g.][and references therein]{2009ApJ...690..216G,2019MNRAS.484.5785G,2009A&A...506L..17L,2013MNRAS.436L.114K,2015ApJ...813L..41A,2016AJ....151...54S,2018A&A...615A.118S,2017ApJ...847....7B,2019MNRAS.487.3990B,2018NatCo...9.4599Z,2020ApJ...896..134P,2020A&A...642A.129S,2021MNRAS.501...50S,2021MNRAS.501.5997T,2022MNRAS.510.3641R,2022MNRAS.513.5238R,2022Natur.609..265J,2023ApJ...943...53K} especially within the past two decades or so.  Additionally, possible QPOs have also been discussed for a few non-blazar AGNs \citep[e.g.][and references therein]{2008Natur.455..369G,2014MNRAS.445L..16A,2015MNRAS.449..467A,2016ApJ...819L..19P,2018A&A...616L...6G,2020ApJ...896..134P} \\

3C 454.3 is one of the brightest and best-studied FSRQs,  with $z = 0.859$ \citep{1989ApJS...69....1H}. It has strong optical polarization, substantial variability throughout the EM spectrum, and non-thermal emission, which are all standard characteristics of an FSRQ  \citep[e.g.][and references therein]{1988ApJ...326L..39S,2006A&A...445L...1F,2006A&A...453..817V,2007A&A...464L...5V,2009A&A...504L...9V,2006A&A...456..911G,2007A&A...473..819R,2008A&A...491..755R,2009ApJ...697L..81B,2009ApJ...699..817A,2010ApJ...712..405V,2010ApJ...715..362J,2010ApJ...721.1383A,2014ApJ...784..141S,2015MNRAS.452.2004M,2017MNRAS.472..788G,2019ApJ...887..185S,2025ApJS..276....1D}. By using optical spectroscopy methods, the mass of the SMBH in 3C 454.3's central engine is estimated to be in the range of (0.5 -- 2.3) $\times$ 10$^{9} \rm{M}_{\odot}$ \citep{2002ApJ...579..530W,2006ApJ...637..669L,2012MNRAS.421.1764S,2017MNRAS.472..788G,2019A&A...631A...4N}. \\

3C 454.3 is among the blazars for which claims of an apparent QPO have been made, in radio, optical, and gamma-ray bands \citep{2021MNRAS.501...50S,2021ApJS..253...10F,2024ApJ...977..166T}. A simultaneous QPO of 47-day period in the gamma-ray as well as optical light curves of 3C 454.3 was reported by \cite{2021MNRAS.501...50S} by analyzing $R$-band optical photometric data taken from 2006 to 2018 from the 70 cm meniscus telescope at Abastumani Observatory, Georgia. In this work, we utilized $\sim$ 19 years of optical $R$-band data of the FSRQ 3C~454.3 to search for QPOs. Because of its brightness and the large number of available observations conducted over an extended temporal span we focused our attention on this source. An initial inspection of the light curve using the Lomb-Scargle Periodogram (LSP) suggested a periodic feature so we then conducted a detailed analysis. In another study conducted by \cite{2021ApJS..253...10F}, the authors reported the detection of three possible periods in the range of $\sim$ 440 to 1100 days. However, their study was not able to fully resolve the temporal evolution of these signals. As the authors noted, the longer-period components appeared fused together in their weighted wavelet Z-transform (WWZ) analysis, which limited their ability to clearly distinguish their individual contributions. As a dense sampling of the source under study is now available with a longer baseline, we have re-analyzed the light curve to extract its frequency and temporal information with appropriate statistical methodologies discussed in the upcoming sections, given the red-noise behavior of the blazar light curves.\\

\begin{table*}[]
    \centering
       \caption{Log of observations obtained from the ARIES 1.3m (A) and 1.04m (B) telescopes.}
    \begin{tabular}{cccccccccc}
    \hline
    Obs date & Telescope & Obs date & Telescope & Obs date & 
    Telescope & Obs date & Telescope & Obs date & Telescope\\
    \hline
    04-11-2020 & A&05-12-2020& A& 10-10-2023& B& 03-11-2023& B& 19-11-2023 & B\\
    05-11-2020 & A&25-10-2021& A& 12-10-2023& B& 05-11-2023& B& 21-11-2023 & B\\
    06-11-2020 & A&27-10-2021& A& 13-10-2023& B& 08-11-2023& B& 09-12-2023 & B\\
    26-11-2020 & A&15-12-2021& A& 15-10-2023& B& 14-11-2023& B& 11-12-2023 & B\\
    27-11-2020 & A&08-10-2023& B& 17-10-2023& B& 17-11-2023& B& 12-12-2023 & B\\
    04-12-2020 & A&09-10-2023& B& 18-10-2023& B& 18-11-2023& B&  & \\
    \hline
    &   
    \end{tabular}
    \label{tab_log}
\end{table*}

In simultaneous observations taken from 1979 to 2013 in five radio frequencies ranging from 4.8 GHz to 37 GHz, a period of $\sim$2000 days in all frequencies was recently reported \citep{2024ApJ...977..166T}. Here, using the optical $R$-band photometric data taken from 2004 to 2023 and detailed photometric results reported in \cite{2025ApJS..276....1D} along with some additional new data taken from two optical telescopes in India, we detected a QPO in 3C~454.3 at a period of $\sim$ 433 days. We explained this QPO by a number of models based on, e.g., binary SMBH systems \citep[e.g.][]{2008A&A...477..407V,2010MNRAS.402.2087V,2015ApJ...813L..41A}, jet-based models, or Lense–Thirring precession of accretion disks \citep[e.g.][]{1998ApJ...492L..59S,2000A&A...360...57R,2004ApJ...615L...5R,2018MNRAS.474L..81L}. \\ 

In \autoref{sec2}, we briefly describe the extensive data set, most of which was discussed in detail in \cite{2025ApJS..276....1D}. We describe the the techniques we employed for the analysis in \autoref{sec3}, while \autoref{sec4} provides a discussion and our conclusions.

\section{DATA ACQUISITION}\label{sec2}
In this work, we have used $R$-band data collected over a period of $\sim$19 years (2004 -- 2023). The majority of data has been provided by the WEBT archive and its collaborators, and includes both published and unpublished data. Apart from that, data from public archives like the Small and Medium Aperture Research Telescope System (SMARTS) and the Steward Observatory have also been used. Details about the telescope facilities utilized can be found in Table 1 of \cite{2025ApJS..276....1D}, where the data reduction is also discussed. We have also included data from two other optical telescopes, 1.30 m and 1.04 m (telescopes A and B from here on, respectively), at the Aryabhatta Research Institute of Observational Sciences (ARIES), India. The total $R$-band light curve is presented in \autoref{R_band_zoomed} with the data log for telescope A and B in \autoref{tab_log}. The highlighted portion in \autoref{R_band_zoomed} is the span during which we find good evidence for a QPO using the techniques discussed in \autoref{sec3}.

\section{Data Analysis Methods and Results}\label{sec3}
We have analyzed those $\sim$ 19 years of R-band optical data and found a fairly strong signal for the presence of QPO from MJD 54980--58450 using three different methods, i.e., Lomb-Scargle Periodogram (LSP), Weighted Wavelet Z-Transform (WWZ) and Phase Dispersion Minimization (PDM). We first applied the LSP to identify a potential periodic component in the light curve. After detecting a candidate period, we used two additional, independent techniques, WWZ and PDM, to corroborate the presence of the QPO. Once all three methods consistently indicated the same periodic signal, we calculated both local and global significance for the detected signal across all methodologies. Descriptions of the LSP, WWZ transform, and PDM methods, along with the estimation of the significance of the detected QPO, are discussed in subsections \ref{LSP}, \ref{WWZ}, \ref{pdm}, and \ref{QPO_sig}, respectively. Least square fitting to the light curve was also performed using an appropriate sinusoidal model as described in \autoref{ls_fit} to get an estimate of the error in the period of the claimed QPO and other associated parameters.

\subsection{Lomb-Scargle periodogram}\label{LSP}
The Lomb-Scargle periodogram \citep{1976Ap&SS..39..447L,1982ApJ...263..835S} is used to detect periodic signals in a data series, especially for unevenly sampled data, which makes it one of the most widely used techniques to search for periodicity. The fundamental machinery behind the LSP is just the Fourier transform, with its power $p$ at frequency $\omega$ given by:
\begin{equation}
    \begin{split}
    p(\omega) = \frac{1}{2}\Big[\frac{[\sum_{i}(m_i -\bar{m}) \cos\omega(t_i -\tau) ]^2}{\sum_i \cos^2\omega(t_i -\tau) } + \\
    \quad \frac{[\sum_{i}(m_i -\bar{m}) \sin\omega(t_i -\tau) ]^2}{\sum_i \sin^2\omega(t_i -\tau) }\Big].
    \end{split}
\end{equation}

Here $m_i$ is the $i$th magnitude value at time $t_i$, $\bar{m}$ is the average magnitude, and the  parameter $\tau$ is given by:
\begin{equation}
    \tan 2\omega\tau = \frac{\sum_i \sin 2\omega t_i}{\sum_i \cos 2\omega t_i} .   
\end{equation}

\begin{figure}
    \centering
    \includegraphics[width=1\linewidth]{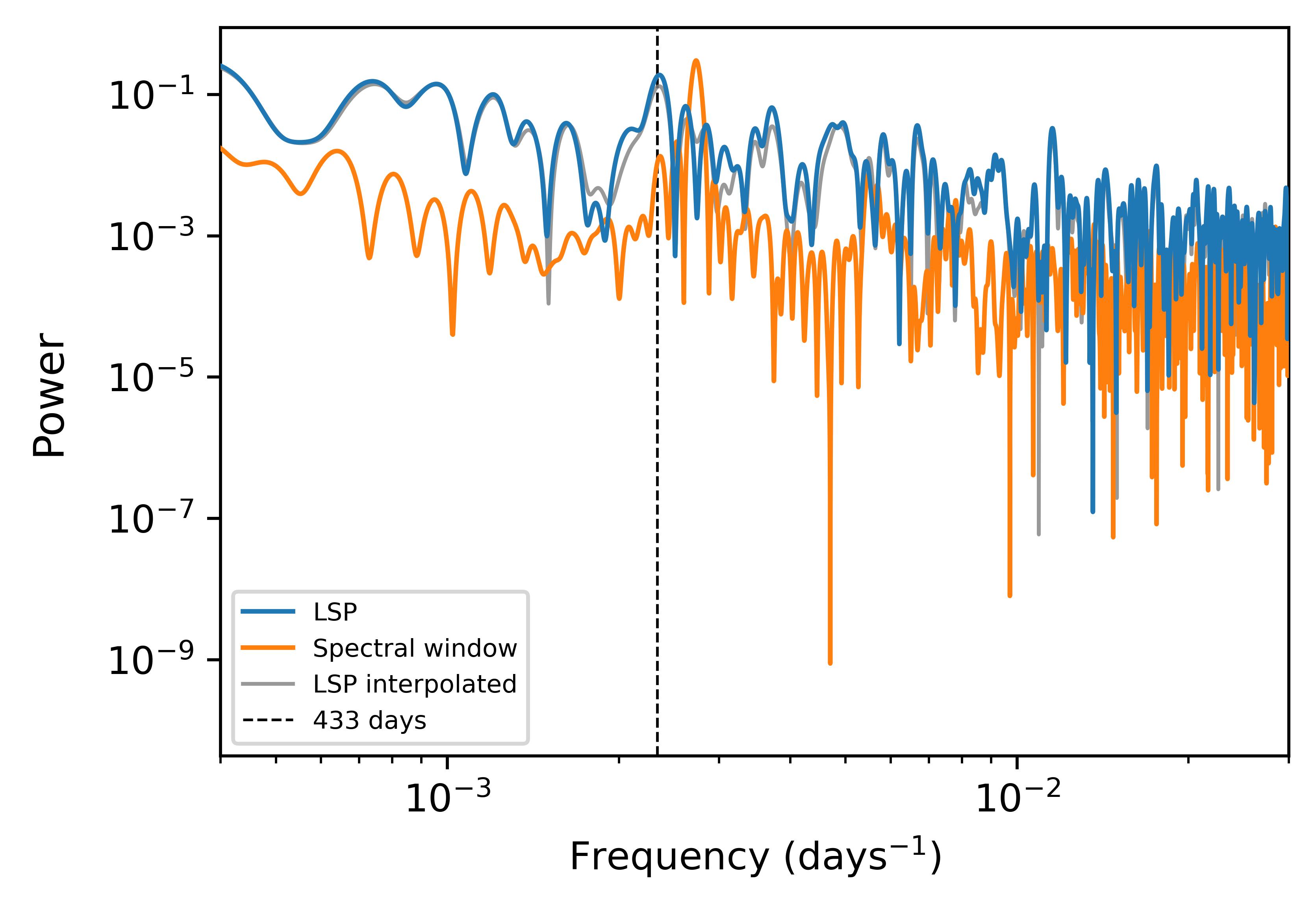}
    \caption{Plot showing the LSP of the original light curve in orange, the spectral window in blue, and the LSP of the gap-interpolated original light curve in grey (see \autoref{gap}).}
    \label{fig:window}
\end{figure}

In this paper, LSP has been implemented using the module {\tt Lomb-Scargle} of the {\tt Astropy\footnote{Astropy is a community-developed Python package that provides tools and resources for astronomy and astrophysics. It is used for a variety of tasks, including astronomical research, data processing, and data analysis.}} package \citep{2022ApJ...935..167A}. The resulting LSP plot is shown in \autoref{fig:window} in orange, with the highest power corresponding to a QPO of $\sim$ 433 days.\\

As shown in \autoref{R_band_zoomed}, several data gaps are present in the light curve, most prominently the usual annual ones, which can lead to aliasing and potentially result in spurious peaks in the LSP. 
When applying LSP to a dataset with gaps, it is vital to take into account its failure modes, i.e., the largest peak in the LSP may not correspond to the true frequency but to some harmonic or alias of it. These effects related to the window function's structure should always be considered. Nevertheless, as mentioned in \cite{2018ApJS..236...16V}, doing so does not guarantee that the true peak is always identified; however, it is still preferable than simply assuming that the highest peak in the periodogram is the correct one. The relevant tests and their results are as follows:\\\\
\indent 1)  \emph{Harmonic test:} We tested whether the 433-day peak could be a harmonic of a longer true period by evaluating the Lomb–Scargle power at the harmonic frequencies $f_{peak}/m$ for at least m $\in$ \{2,3\}, corresponding to $\approx$ 866 days and 1299 days. As no significant peak compared to 433 days was found at either of these periods, we conclude that the 433-day peak is not a harmonic.\\

\indent 2) \emph{Aliasing test:} To test whether the 433-day peak originated from window-function aliasing, we used the strongest window-function component at 365 days ($\delta f=1/365 \ {\rm day}^{-1}$) and searched for periodogram peaks at alias frequencies $\lvert \ f_{peak} \pm n\delta f \ \rvert$ for at least $n\in$ \{1,2\} corresponding to $\approx$ 198 days, 128 days, 315 days and, 2324 days. None of the alias candidates produced Lomb–Scargle power greater than or equal to the 433-day peak, indicating that the 433-day QPO is not a product of aliasing.\\

A recent study \citep{2025MNRAS.tmp.1751P} demonstrated that the reliability of period detection deteriorates significantly when gaps account for more than 50\% of a light curve. In our case, the fraction of missing data is approximately 37\%, which remains below this threshold. Therefore, the gaps in our dataset are not expected to introduce the severe degradation reported in that study. We address additional aspects of the issues arising from data gaps and possible solutions to them in \autoref{gap}, where interpolation and its use in distinguishing actual QPOs from yearly seasonal gap artifacts using simulated light curves are discussed. We simulated light curves for spectral indices ranging from $\alpha = 1.50$ to $2.50$, incorporating QPO periods between 250 and 500 days (see \autoref{fig_A1}). We found that for more than $98\%$ of the tested periods in the simulations, the change in QPO power before and after interpolation remains below $15\%$ across all spectral indices. Regarding global significance (see \autoref{QPO_sig}), the original light curve exhibited a change of less than $5\%$. In contrast, for simulated cases where the maximum power change occurred, the corresponding change in global significance reaches approximately $17\%$. Additional details and extended results are presented in \autoref{gap}.\\

The LSP method is robust and provides a good estimate of the overall power of the frequency components present, but it does not describe when, during the time series, a particular signal was present (at least for a quasi-periodic signal), and for how long it persisted. To obtain information of this type, an alternative approach is necessary, as discussed in \autoref{WWZ}.

\begin{figure*}
    \centering
    \includegraphics[width=16cm, height=8.5cm]{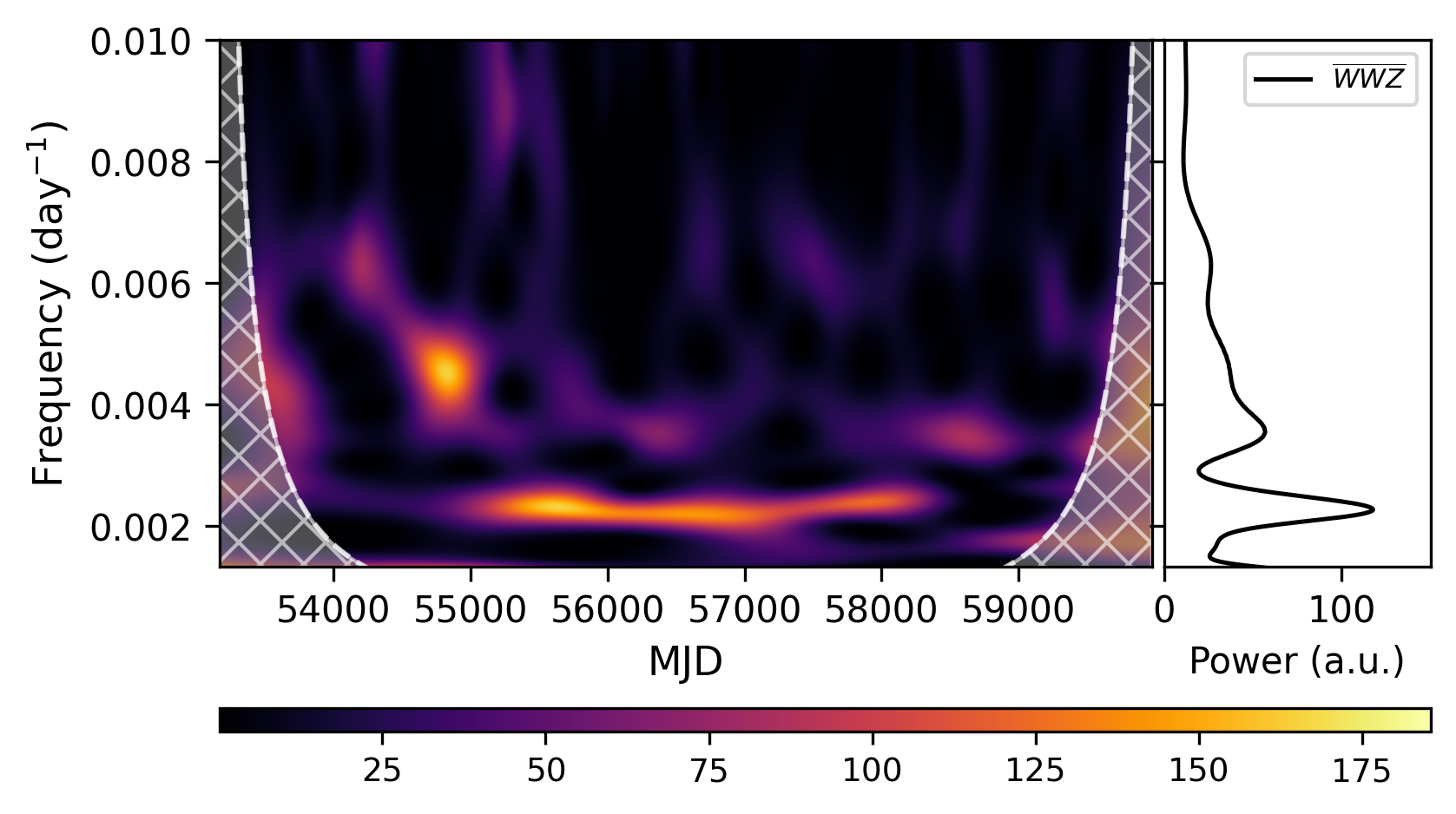}
    \caption{On left: Result of the WWZ analysis with a dominant signal at 0.00231 day$^{-1}$ from MJD 54980 to 58450. The white dashed line separates the region in the WWZ plot, where edge effects become dominant, with our QPO of $\sim$ 433 days in the safe region. On right: Time-averaged WWZ ($\overline{WWZ}$) plot.}
    \label{wwz_plot}
\end{figure*}

\subsection{Weighted wavelet Z-transform}\label{WWZ}
Another technique for detecting periodic signals in unevenly sampled data that is employed frequently is the WWZ transform \citep[e.g.][]{1996AJ....112.1709F}.  
This technique not only detects periodicities but also tracks the evolution of period, amplitude, and phase in the time series under study, which makes it a great choice when searching for QPOs. Due to its unique property of exploring both time and frequency domains, unlike periodograms, it generates results as a function of both frequency and time. This yields a 3-D plot with time normally shown along the X-axis, the frequency along the Y-axis, and the Z-axis representing the WWZ output, which is typically plotted using a color bar. The wavelet transform (WT) of a function $y(t)$ is given as:
\begin{equation}
    \label{WT}
    \begin{split}
    WT(\omega,\tau;y(t)) &=  \omega^{1/2}  \int y(t)f^*(\omega(t-\tau)) \,dt \\
     & = \omega^{-1/2} \int y(\omega^{-1}z + \tau)f^*(z) \,dz. \\
    \end{split}
\end{equation}

The function $f(z)$ in \autoref{WT} is the wavelet kernel, also known as the mother wavelet, with $f^*(z)$ being its complex conjugate. As can be seen, the wavelet transform depends on two factors: the frequency $\omega$ and the time shift $\tau$. We have used the frequently employed Morlet wavelet, which has a Gaussian decay profile of the following mathematical form:\\
\begin{equation}
    \begin{split}
        f(z) &= e^{-cz^2}(e^{iz}-e^{-1/4c}), \\
        & = e^{-c\omega^2(t-\tau)^2}(e^{i\omega(t-\tau)}-e^{-1/4c}). \\
    \end{split}
    \label{wwz_eq}
\end{equation}
Here, the constant $c$ controls the rate at which the chosen wavelet decays. It is set in such a way that the exponential term in \autoref{wwz_eq} decays significantly in one cycle. A critical reason for having the term $e^{-1/4c}$ in the above equation is to make the average value of the wavelet zero, such that:
\begin{equation}
    \int_{-\infty}^{+\infty} f(z) \,dz = 0 .\\
\end{equation}

Although $c$ can be treated as a free parameter, a popular choice for this constant is $1/8\pi^2$, the value we have adopted in this analysis. That makes the second terms within the parentheses in \autoref{wwz_eq}, i.e., $e^{-1/4c},$ negligible, and hence gives us the so-called abbreviated Morlet transform:
\begin{equation}
    f(z) = e^{iz - cz^2}.
\end{equation}

To perform the WWZ transform, a Python package, {\tt wwz\footnote{\url{https://github.com/skiehl/wwz}}} \citep{2023ascl.soft10003K}, based on Foster's algorithm, has been used. The resulting WWZ plot can be seen in \autoref{wwz_plot} with a prominent quasi-periodic signal of 0.00231 day$^{-1}$ or $\sim$ 433 days persisting for around 3400 days with some modulation in frequency that coincides with the LSP peak at $\sim$ 433 days in \autoref{fig:window}. The cross-hatched region bounded by white dashed lines demarks the cone of influence (COI). The construction of the COI in WWZ analysis is crucial when signals appear near the edges of the time domain. In such cases, edge effects can artificially stretch the signal, creating the impression that a periodic component lasts longer than it actually does. This effect is particularly exaggerated at lower frequencies, where the wavelet extends over a larger portion of the time series. Unless a sufficiently large data set is available so that the edge effects are minimal, the COI should always be defined. Omitting it can lead to misleading conclusions about the duration of the detected periodicity. As can be seen in \autoref{wwz_plot}, the 433-day QPO is well within the safe region.

\subsection{Phase dispersion minimization}\label{pdm}

In order to increase one's confidence in a detected signal, another technique known as phase dispersion minimization (PSD) \citep{1978ApJ...224..953S} can be performed, which is well-suited for searching non-sinusoidal signals. A light curve with $N$ number of ($t_i, m_i$) sets, where $m_i$ is the magnitude or flux at time $t_i$, is phase folded at a trial period and is then divided into $M$ phase bins, with each bin having $n_j$ data points with similar phases. The variance in each bin is given by:\\
\begin{equation}
    s_j^2 = \frac{\sum_{j=1}^{M} (x_{kj}-x_j)^2}{n_j -1} ,
\end{equation}
where $x_{kj}$ is the $k$th data point in the $j$th phase bin. Then the net variance for all the phase bins at a particular trial period is given by:\\
\begin{equation}
    s^2 = \frac{\sum_{j=1}^{M} (n_{j}-1)s_j^2}{\sum_{j=1}^{M} n_j -M}.
\end{equation}

If the variance of the full light curve is given by:\\
\begin{equation}
    \sigma^2 = \frac{\sum_{i=1}^{N} (m_{i}-\bar m)^2}{N -1},
\end{equation}
then the PDM statistic, $\theta$, is the ratio of the net variance of the phase bins at a trial period to the variance of the full light curve, i.e., $\theta$ = ${s^2}/{\sigma^2}$. Thus  $\theta$ is essentially a measure of the scatter of sample variance around the mean of the light curve.\\

\begin{figure}
    \centering
    \includegraphics[width=8.5cm, height=5.2cm]{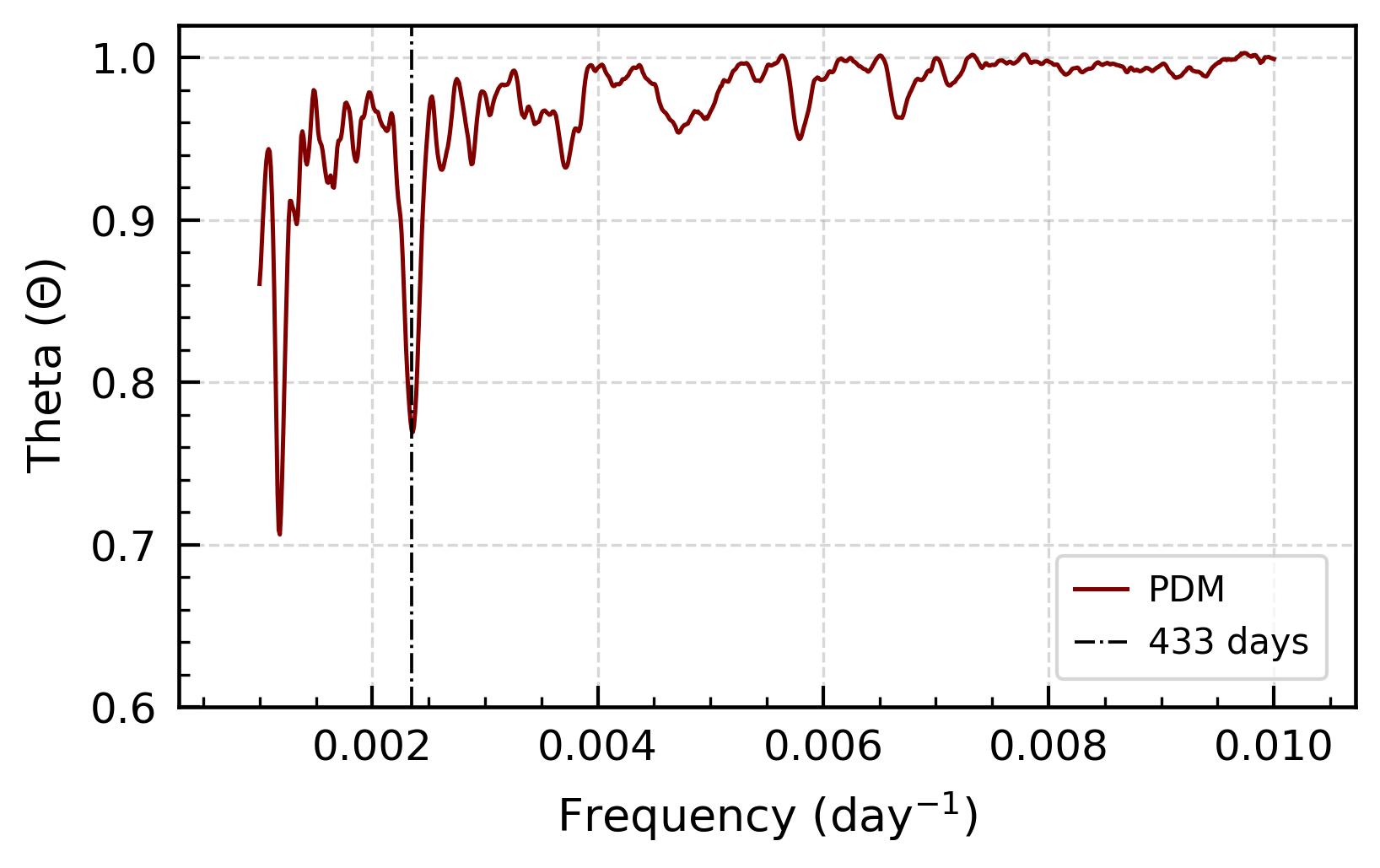}
    \label{fig:subfig3}
    \caption{PDM plot and 433 days QPO dip shown by dot-dashed line, which is also detected in the LSP and time-averaged WWZ.}
    \label{pdm_plot}
\end{figure}

To perform the PDM analysis, we used the module {\tt PyPDM} from the {\tt PyAstronomy\footnote{PyAstronomy (PyA) is a collection of astronomy-related packages hosted on GitHub.}} \citep{pya} package.
For a non-sinusoidal or aperiodic variation, $s^2 \approx \sigma^2$, and $\theta$ is close to 1, whereas if a periodic component is present in the light curve, $\theta$ is expected to be low, so the plot will have a local minimum at that period. So, by repeating this process for a set of trial periods, one can identify any dominant period(s) in the time series. As can be seen in \autoref{pdm_plot}, a dip is detected in the PDM plot for 3C 454.3 corresponding to a 433-day period, as detected earlier in both LSP and WWZ analyses. Once a putative QPO signal has been detected, its significance must be estimated, as discussed in \autoref{QPO_sig}.

\begin{figure}
    \centering
    \includegraphics[width=1.0\linewidth]{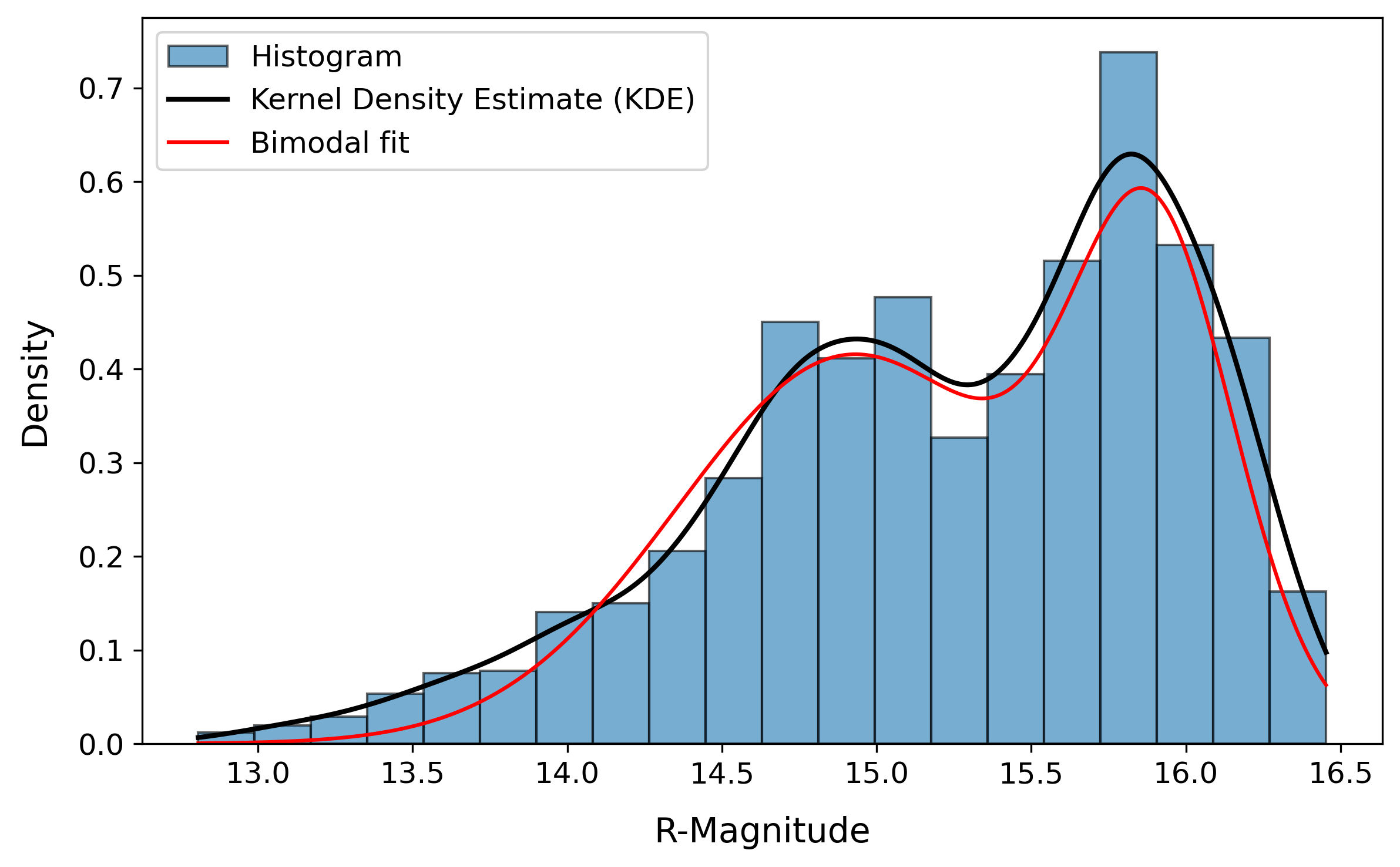}
    \caption{Probability distribution function for the R-band optical light curve. To estimate the function, KDE was performed (shown in black), revealing a bimodal distribution, which was then fitted using a bimodal function (shown in red). }
    \label{fig:pdf}
\end{figure}

\begin{figure*}
    \centering
    \includegraphics[height=13.5cm, width=13.5cm]{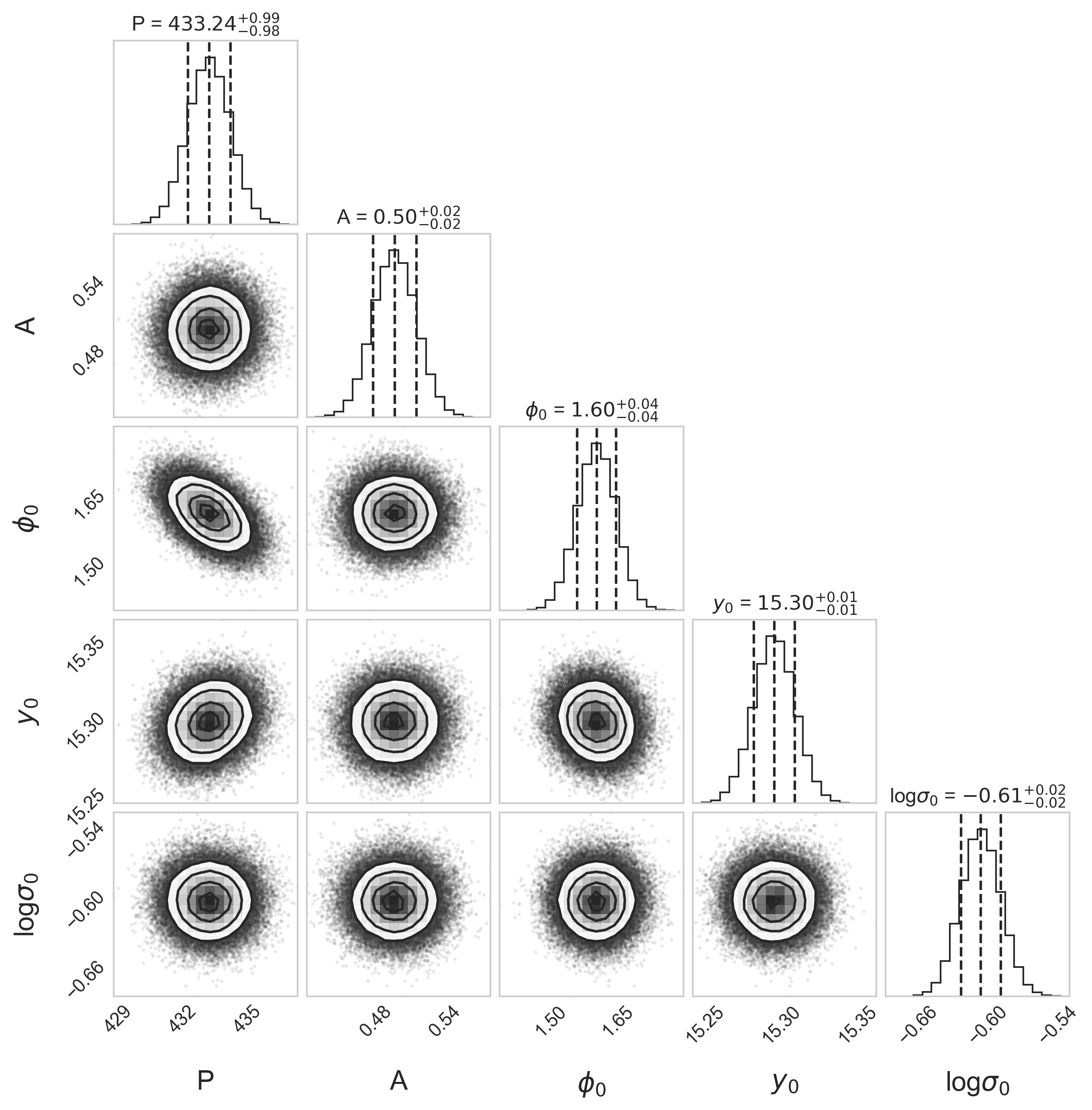}
    \caption{Corner plot showing a two-dimensional projection of the posterior probability distribution of the sinusoidal model parameters given in \autoref{sine}.}
    \label{fig_mcmc}
\end{figure*}

\subsection{QPO significance estimation}\label{QPO_sig}
As the blazar light curves are red-noise dominated, the significance of a probable signal in LSP and PDM needs to be estimated. To do so, the underlying models are assumed, giving rise to the observed power spectrum density (PSD) and the probability distribution function (PDF) of the optical light curve. The best model was chosen such that Akaike's Information Criteria, $AIC = 2k - 2 ln(L)$, is minimized, where $k$ is the number of free parameters and $ln(L)$ is the log-likelihood. For modeling the PSD, three models, namely power-law \citep{2004MNRAS.348..783M, 2010ApJ...722..520A, 2020ApJ...891..120B}, broken power-law \citep{2024A&A...686A.228O}, and bending power-law \citep{2016MNRAS.461.3145V}, were chosen because of the essentially red-noise behavior of the underlying PSD. Among these, the broken power-law had the lowest AIC value of 91797, while the power law and smoothly bending power law returned the AIC values of 101944 and 91882, respectively. In analyzing the PDF we first performed an Anderson-Darling test, which rejected the null hypothesis that the PDF distribution of the light curve is Gaussian at a significance level of $10^{-3}$. To get an idea of the shape of the PDF we performed a kernel density estimation (KDE), which revealed a bimodal distribution as shown in \autoref{fig:pdf}. We then made a bimodal fit to the PDF and computed its AIC, which came out to be 4583, which we compared with the AIC of the Gaussian fit, which was 4987. Thus we prefer the bimodal function. The fitting functions and their corresponding fitting parameters and AIC values can be found in \autoref{tab_fit}.\\

The final PSD and PDF model functional forms are given below in \autoref{eq_BPL} and \autoref{eq_bimodal}.  
\begin{equation}
    PSD(f) =
\begin{cases}
A \left( \dfrac{f}{f_b} \right)^{-\alpha_{\text{low}}}, & \text{if } f \leq f_b \\
A \left( \dfrac{f}{f_b} \right)^{-\alpha_{\text{high}}}, & \text{if } f > f_b .
\end{cases}
\label{eq_BPL}
\end{equation}
Here $A$ is the amplitude, $f_b$ is the break frequency, and $\alpha_{low}$ and $\alpha_{high}$ are the power-law indices below and above $f_b$, respectively.

\begin{equation}
\begin{split}
PDF(m) =\ & A_1 \exp\left( -\frac{(m - \mu_1)^2}{2\sigma_1^2} \right) \\
       & + A_2 \exp\left( -\frac{(m - \mu_2)^2}{2\sigma_2^2} \right)
\end{split}
\label{eq_bimodal}
\end{equation}
Here $A_i$, $\mu_i$, and $\sigma_i$ represent the amplitude, mean, and standard deviation of a particular Gaussian, respectively.\\

Once the appropriate parameters for the underlying PSD and PDF were extracted, $2 \times 10^6$ light curves were simulated, with the same data sampling as the original light curve, using the EMP13 technique \citep{2013MNRAS.433..907E} implemented in the Python code {\tt lcsim\footnote{https://github.com/skiehl/lcsim}} \citep{2023ascl.soft10002K}. Unlike the TK95 algorithm \citep{1995A&A...300..707T}, in which only the PSD, along with the mean and standard deviation of the original light curve, is preserved in the simulated light curves, EMP13 also preserves the PDF of the light curve.
Using these simulated LCs, we calculated the global significance of the detected period, which incorporates the effect of the presence of such a dominant signal at other frequencies, as opposed to the local significance (also called the single period p-value) found only at a target frequency. To evaluate the global significance, we accounted for the \emph{``look-elsewhere''} effect by estimating the probability of obtaining apparently significant peaks purely from noise when scanning across a broad frequency range as described in \cite{2022ApJ...926L..35O}. To do so, we first calculated the p-value corresponding to the 433-day QPO peak, $p_{qpo}$, which corresponds to the local significance, by finding the number of simulations that resulted in a peak at 433 days with equal or more power, and then dividing this number by the total number of simulations. We found a local p-value of $4.71 \times 10^{-6}$, corresponding to a local significance of $4.43\sigma$. Then we find the LSP for each simulation, and the strongest peak and its corresponding period are noted. Next, following the above procedure for the local \emph{p-value}, we compute $p_{sim}$. Finally, we count all the simulations whose $p_{sim} \leq p_{qpo}$ and divide this number by the total number of simulations to get the global \emph{p-value} and the corresponding significance. We found a global \emph{p-value} of $1.87 \times 10^{-3}$ corresponding to a global significance of 2.90$\sigma$. 

We also calculated the global p-values for the WWZ and PDM methods, which were similar to, but greater than, the global p-value for the LSP approach, at $2.21 \times 10^{-3}$ (WWZ) and $2.21 \times 10^{-3}$ (PDM). As we have employed three different techniques, among which the LSP returned the smallest global p-value (i.e., the highest significance), we now apply the Bonferroni correction \citep{abdi2007bonferroni} to obtain a single global p-value across all methodologies. This correction is applied when multiple independent tests are performed on a dataset, thereby increasing the chance of at least one false positive. Though more conservative, the Bonferroni correction addresses this by tightening the threshold. If $n$ represents the number of tests performed and $p$ represents the smallest p-value among the tests, then the corrected global p-value is given by $n\times p$. We found that the global p-value across all methods is $5.61 \times 10^{-3}$, corresponding to $2.53\sigma$.

\subsection{Least square fitting of the light curve}\label{ls_fit}
We fitted a sinusoidal model to the light curve of the form:
\begin{equation}
    y_{model}(t) = A\sin(2\pi f(t_i-t_0) -\phi_0) \ + \ y_0
    \label{sine}
\end{equation}
Where $A$ is the amplitude, $t_0 = 57000$ (mid-point of the light curve), $P=1/f$ is the period, $\phi_0$ is the phase of the sinusoid, and $y_0$ is the mean of the light curve. To find the best parameters, we maximized the following likelihood:\\
\begin{equation}
    ln \ L = -\frac{1}{2}\sum_{i}^{N}\left( \frac{(y_i  -  y_{model})^2}{\sigma_i^2  +  \sigma_0^2} + ln(\sigma_i^2 +\sigma_0^2)  \right)
\end{equation}
Here $y_i, \ \sigma_i,\ and \ \sigma_0$ are the light curve data points, observational errors, and Gaussian white noise, respectively. Using the maximum likelihood estimate (MLE), we found the best-fitting parameters. As the MLE does not directly provide parameter errors, we used the {\tt emcee\footnote{https://emcee.readthedocs.io/en/stable/}} \citep{2013PASP..125..306F} package to perform MCMC sampling, initialized with the MLE-derived estimates. We reported uncertainties as the $68\%$ confidence intervals derived from the marginalized posterior distribution, as shown in \autoref{fig_mcmc}. We found that a period of $\approx$ 433 days fits the light curve well. A general practice is to fit the LSP peak to quote the error in period or frequency, but that would not be a correct approach, as the number of data points and signal-to-noise ratio do not affect the peak width but only the peak height, as explained in section 7.4 of \cite{2018ApJS..236...16V} which makes the MCMC fitting a better approach.

\begin{table}[]
    \centering
    \caption{Fitting functions and their associated parameters.}
    \begin{tabular}{llc}
    \hline
    Fitting function & Fitting parameters& AIC\\
    \hline
          & A = 24.223 $\pm$ 0.943& \\
     Broken power-law  &  $f_b$ = 0.0024 $\pm$ 0.0001& 91797.28\\
       function   &  $\alpha_{low}$ = 0.553 $\pm$ 0.013& \\
          &  $\alpha_{high}$ = 1.442 $\pm$ 0.044& \\
          \hline
          & $A_1$ =  0.516 $\pm$ 0.022& \\
          & $\mu_1$ = 15.345 $\pm$ 0.040& \\
Bimodal  & $\sigma_1$ = 0.736 $\pm$ 0.038& 4582.92\\
function         & $A_2$ = 0.641 $\pm$ 0.166& \\
         & $\mu_2$ = 15.345 $\pm$ 0.021& \\
         & $\sigma_2$ = 0.068 $\pm$ 0.021& \\
        \hline
    \end{tabular}
    \label{tab_fit}
\end{table}

\section{Discussion and conclusions}\label{sec4}
This project involved data from several telescopes worldwide in collaboration with the Whole Earth Blazar Telescope (WEBT) team. We evaluated the Lomb-Scargle periodogram, the weighted wavelet Z-transform, and the phase dispersion minimization method to search for QPOs. We have detected a signal in the light curve segment highlighted in \autoref{R_band_zoomed} from MJD 54980--58450 that has a period of $\sim$ 433 days. This corresponds to $\approx$ 8 cycles, as can be seen in \autoref{wwz_plot}. This signal is detected at a global significance of $2.53\sigma$. \cite{2021ApJS..253...10F} reported the possible presence of QPOs with data spanning from 2006--2018. In our re-analysis of the light curve, we identify only a 433-day periodicity in the light curve. The difference in results can arise because our dataset contains a larger number of observations with 12096 data points (compared to 8523 points in the study mentioned above) over the same interval, and also spans a longer temporal baseline. These improvements reduce edge effects and enhance the reliability of signal detection. Our results indicate that a 433-day signal is consistently present from 2009 to 2018, and its statistical significance has been rigorously assessed using the methods described in this work.\\

Quasi-periodic oscillations in blazars are generally interpreted within three broad frameworks: models based on jet dynamics, those linked to processes around the central supermassive black hole (SMBH), and scenarios involving binary SMBH systems. As the object under study is a blazar, with its jet pointing close to the line of sight of the observer, the jet emission is Doppler-boosted; hence, any variability and, in turn, any QPOs detected probably can be explained using jet emission models. \cite{1985ApJ...298..301H} proposed shocks as the main driving component behind the outbursts observed in BL Lac in 1982 and 1983, causing the emission, if effect, to be polarized due to compression of initially random magnetic field. So, when a shock propagates down the jet in a region with a dominant helical magnetic field \citep[e.g.][]{2008Natur.452..966M}, it can produce quasi-periodic oscillations due to the delicate dependence of the Doppler factor on the viewing angle, $\theta(t)$, given that the shock speed changes between the helical winds \citep{2011JApA...32..147W}, making the effect similar to a jet changing its direction. The observed flux $F_{\nu}$ is given by:\\
\begin{equation}
    F_{\nu} = \delta(t)^{3+\alpha} F^{'}_{\nu^{'}} ,
\end{equation}
where $F^{'}_{\nu^{'}}$ is the rest frame flux and the Doppler factor, $\delta(t)= 1/\Gamma (1-\beta \cos \theta(t))$. For a blazar with optical spectral index, of say, $\alpha= 1.5$, and viewing angle in the range $1^{\circ}-5^{\circ}$, a change in viewing angle even by $1^{\circ}$ would cause the apparent flux to change by a factor of two or more \citep[e.g.][]{2020ApJ...891..120B}.

Another possible way in which a QPO involving a propagating \citep{1992vob..conf...85M} or standing \citep{2008Natur.452..966M} shock could arise would be from the presence of a dominant cell in the strong turbulent flow behind these shocks in the jet. Because of the stochastic nature of the turbulence, the QPO would typically appear only for a limited time \citep[e.g.][]{2011JApA...32..147W}. 

Nevertheless, emission variations produced in a binary super-massive black hole (SMBH) system and in an SMBH system can provide viable explanations for such a QPO  in 3C 454.3, as it is an FSRQ, having a bright AD that leaves its imprint visible in the spectral energy distribution (SED)\citep{2011MNRAS.410..368B}\\

Another possible mechanism involves a tilted AD and the Lense-Thirring (LT) precession \citep{1984GReGr..16..711M}, occurring due to relativistic frame dragging by the central SMBH due to its spin.
The precession time period ($T_{LT}$) around a Kerr BH with spin parameter $a_s$ and mass $M$ at a distance $r$ in the equatorial plane of the BH is given by:\\
\begin{equation}
    T_{LT} = 0.18\left(\frac{1}{a_s}\right) \left(\frac{M}{10^9 M\odot}\right) \left( \frac{r}{r_g} \right)^3 \text{days}
\end{equation}

Considering a maximally rotating BH, this would happen around the inner part of the AD corresponding to $\sim 10$ Schwarzschild radii ($r_g$), for the detected period.
\cite{2018MNRAS.474L..81L} conducted a study using 3-D general relativistic magnetohydrodynamical simulations (for different magnetic fluxes and resolutions) running for $(0.3-1.2) \times 10^5\ t_g$. They concluded that the disc-jet system undergoes LT precession, aligning with the BH spin, with faster alignment for stronger magnetic flux strengths. This would result in jet precession, which would, in turn, cause the Doppler factor, hence observed brightness, to vary quasi-periodically. \\

Apart from jet and AD based models, binary SMBH systems can also produce quasi-periodic oscillations either directly, if the secondary SMBH impinges onto the AD of the primary \citep[e.g.][]{1996ApJ...460..207L, 2023ApJ...957L..11G} or by inducing periodic fluctuations in the accretion rate because of an elliptical orbit \citep[e.g.][]{2022ApJ...929..130W}. Due to the cosmological redshift, the period in the source frame is shortened by a factor of (1+z), or $\sim$ 233 days for a period of $\sim$ 433 days in the observer frame. If we consider this to be the orbital period $(T)$ of the secondary, with mass $(m)$, around the central SMBH, having mass $(M)$ and assuming a Keplerian orbit, the time period is given by:\\
\begin{equation}
    T = 1.08 \times 10^6 \left(\frac{a}{1\text{ pc}}\right)^{3/2}\left(\frac{M+m}{10^9M_{\odot}}\right)^{-1/2} \text{days}
\end{equation}
Taking the central SMBH mass to be $\sim0.85$ $\times$ $10^9$ $M_{\odot}$ \citep{2019A&A...631A...4N}, we find that the binary SMBH system will have a separation, $a$, of order of a few milliparsec. The gravitational wave-driven timescale ($t_{GW}$) \citep{1964PhDT........51P} at such separations will be $\sim 10^4 -10^6$ years, assuming a mass ratio between 0.1 and 0.001 and a circular orbit. \\

Geometrical models can explain quasi-periodic variations without requiring a change in the intrinsic jet flow owing to the changing Doppler boosting. These models include jet precession \citep{2013MNRAS.428..280C}, a helical jet \citep{1993ApJ...411...89C, 2018NatCo...9.4599Z}, or a curved helical jet \citep{2015MNRAS.454..353R, 2021MNRAS.501...50S}. These phenomena take place especially in binary SMBH systems, as is detailed in the models for the blazar OJ 287 (see \cite{2024RNAAS...8..276V} and \cite{2024ApJ...968L..17V}  and references therein). In this case, the origin of each jet is in constant motion, which causes it to change direction with respect to the observer in a cyclic manner due to aberration. For equal mass binaries in a circular orbit, the pattern is simplest: each jet may point toward us (or closest to our direction) once per orbit cycle. This may produce two equally spaced radiation peaks per orbital cycle. The peaks from the second black hole's jet lie halfway between the peaks from the first jet only when the black hole masses are equal and the orbit is circular; otherwise, we may have two sets of brightness peaks that are arbitrarily displaced relative to each other but still have the same periodicity, though different amplitudes. 
The precession of the binary's major axis complicates this simple picture; as a result, we do not see exact periods but rather quasi-periodicity in the light curve. One example of this kind is the beaming model by \citet{1998MNRAS.293L..13V}, put forth to explain the quasi-periodicity of the double-peaked optical outbursts observed in OJ~287. Spin-orbit interaction has a qualitatively similar effect. The binary motion, therefore, produces a helical jet structure, meaning that the times of the maximum radiation peaks depend on the distance along the jet, which typically varies with the observing frequency. In the case of 3C 454.3, the 433-day quasi-periodicity appears occasionally as an exact interval between two peaks in the light curve, e.g., the peak at MJD 55530 and the 3 peaks before and after it at this time interval. Unlike in the OJ 287 binary model, where orbital parameters are claimed to be fully determined \citep{2018ApJ...866...11D} and lead to predictable flares, here we have too many parameters to determine all at once. Trying to do so would be an extremely interesting future study. Since the period is relatively short in comparison with that of OJ 287, one could perhaps reach the level of predictability in 3C 454.3 after a few more orbital cycles if this FSRQ indeed houses a binary SMBH system.\\

There have been instances in which periodicity in an SMBH system evolves, appears, or disappears, as in our case. An illustrative example is the blazar PKS 2131-021, where a periodic signal was detected in radio light curves from UMRAO, OVRO, and Haystack Observatory \citep{2022ApJ...926L..35O, kiehlmann2025}. When the data were divided into three epochs, the authors found periodicity only in the first and third epochs; sinusoidal variations were absent in epoch 2 (1983–2003). They suggest that because blazar variability is largely stochastic, sinusoidal or quasi-periodic variations may sometimes represent a special case even in binary SMBH systems. According to \cite{2022ApJ...926L..35O}, the observed variability can be interpreted as a superposition of two processes: (1) an intermittent periodic component from binary SMBH interactions, and (2) stochastic jet emission from multiple regions. Variations in their relative strengths naturally lead to different observed behaviors from noisy quasi-periodic signals, from cleaner periodic modulations when the binary dominates, to more complex, non-sinusoidal structures when stochastic emission is dominant. Although a candidate case, it still provides insight into how periodic variation may appear intermittently.\\

However, it must be noted that \citet{2018MNRAS.481L..74H} constructed mock blazar populations based on the luminosity functions of BL Lacertae objects and flat-spectrum radio quasars with $z\leq2$ to model the stochastic gravitational-wave background from supermassive black hole binaries. Comparing their predictions to pulsar timing array limits, they found consistency only if $\sim 0.1\%$ of blazars host a binary with an orbital period $<5$ years. A recent update using 15 years of NANOGrav data raises this limit to $\lesssim 0.5\%$ \citep{kiehlmann2025}. Hence, we do not expect the SMBH binary explanation for the blazar QPO phenomenon to be correct very often. Apart from that, the inferred GW-driven timescale ($t_{GW}$) also makes this scenario statistically less plausible.\\ 

Although the periodic signal reaches a global significance of $\approx 2.5\sigma$, we note that stochastic variability in blazar light curves can occasionally produce QPO-like features over finite time intervals. Red-noise processes are known to generate spurious peaks in periodograms with moderate statistical significance. While our simulations account for this effect, we cannot entirely exclude the possibility that the feature represents a statistical fluctuation of an underlying stochastic process. Therefore, the detected period should be interpreted conservatively.\\

In summary, while binary SMBH scenarios provide an appealing framework to explain quasi-periodic oscillations in blazars, current observational constraints make them unlikely to be generally applicable. Models relying on processes within relativistic jets, conceivably initiated in the accretion disks around a single SMBH, offer more plausible explanations for the observed QPO signatures. Currently, the data do not permit a definitive distinction among these scenarios; however, future long-term monitoring may help establish the physical origin of the apparent QPO and determine whether it reflects a persistent dynamical mechanism or a transient manifestation of stochastic variability.

\software{Astropy, wwz, PyAstronomy, lcsim, emcee}
\facilities{ARIES Observatory, SMARTS Observatory, Steward Observatory, and WEBT Collaboration (\cite{2025ApJS..276....1D}) }

\section*{Acknowledgements}
We thank the reviewer for their useful comments, which helped us improve the manuscript.

Based on data taken and assembled by the WEBT collaboration and stored in the WEBT archive at the Osservatorio Astrofisico di Torino – INAF (\url{https://www.oato.inaf.it/blazars/webt/}). This study was based in part on observations conducted using the 1.8m Perkins Telescope Observatory (PTO) in Arizona, which is owned and operated by Boston University. This paper has made use of up-to-date SMARTS optical/near-infrared light curves that are available at {\url{https://www.astro.yale.edu/smarts/glast/home.php}}. Data from the Steward Observatory spectropolarimetric monitoring project were used. This program is supported by Fermi Guest Investigator grants NNX08AW56G, NNX09AU10G, NNX12AO93G, and NNX15AU81G. The ZTF survey is supported by the U.S. National Science Foundation under grants no. AST-1440341 and AST-2034437. The Skinakas Observatory is a collaborative project of the University of Crete, the Foundation for Research and Technology -- Hellas, and the Max-Planck-Institut f$\ddot{u}$r Extraterrestrische Physik. This article is partly based on observations made with the IAC80, the STELLA, and the LCOGT 0.4 m telescopes. The IAC80 telescope is operated by the Instituto de Astrof\'{i}sica de Canarias in the Spanish Observatorio del Teide on the island of Tenerife. Many thanks are due to the IAC support astronomers and telescope operators for supporting the observations at the IAC80 telescope. The STELLA robotic telescopes are an AIP facility jointly operated by AIP and IAC. One of the nodes of the LCOGT 0.4m telescope network is located in the Spanish Observatorio del Teide. The R-band photometric data from the University of Athens Observatory (UOAO) were obtained after utilizing the robotic and remotely controlled instruments at the facilities. 

ACG is partially supported by the CAS President’s International Fellowship Initiative (PIFI), Grant No.\ 2026PVA0040. The Abastumani team acknowledges financial support by the Shota Rustaveli NSF of Georgia under contract FR-24-515.
CMR, MV, and MIC acknowledge financial support from the INAF Fundamental Research Funding Call 2023 (PI: Raiteri). The research at Boston University was supported in part by several NASA Fermi Guest Investigator grants; the latest are 80NSSC22K1571 and 80NSSC23K1507. RB, ES, and AS were partially supported by the Bulgarian National Science Fund of the Ministry of Education and Science under grants KP-06-H68/4 (2022), KP-06-H88/4 (2024), and KP-06-KITAJ/12 (2024).  GD, OV, MDJ, and MS acknowledge support from the Astronomical station Vidojevica, funding from the Ministry of Science, Technological Development and Innovation of the Republic of Serbia (MSTDIRS, contract no.\ 451-03-136/2025-03/200002), by the EC through project BELISSIMA (call FP7-REGPOT-2010-5,  No.\ 256772), the observing and financial grant support from the Institute of Astronomy and Rozhen NAO BAS through the bilateral SANU-BAN joint research project ``GAIA astrometry and fast variable astronomical objects", and support by the SANU project F-187. Also, this research was supported by the Science Fund of the Republic of Serbia, grant no.\ 6775, Urban Observatory of Belgrade - UrbObsBel. The NRIAG team acknowledges financial support from the Egyptian Science, Technology \& Innovation Funding Authority (STDF) under grant number 45779. The R-band photometric data from the University of Athens Observatory (UOAO) were obtained in the frame of {\it BOSS} Project \citep{2019Galax...7...58G} after utilizing the robotic and remotely controlled instruments at the University of Athens \citep{2016RMxAC..48...22G}. The IAA-CSIC co-authors acknowledge financial support from the Spanish ``Ministerio de Ciencia e Innovaci\'{o}n" (MCIN/AEI/ 10.13039/501100011033) through the Center of Excellence Severo Ochoa award for the Instituto de Astrof\'{i}sica de Andaluc\'{i}a-CSIC (CEX2021-001131-S), and through grants PID2019-107847RB-C44 and PID2022-139117NB-C44. Some of the data are based on observations collected at the Observatorio de Sierra Nevada, which is owned and operated by the Instituto de Astrof\'{i}sica de Andaluc\'ia (IAA-CSIC), and at the Centro Astron\'{o}mico Hispano en Andalucía (CAHA); which is operated jointly by Junta de Andaluc\'{i}a and Consejo Superior de Investigaciones Cient\'{i}ficas (IAA-CSIC). LC acknowledges the support from the Tianshan Talent Training Program (grant No. 2023TSYCCX0099). JHF's work is partially supported by the National Natural Science Foundation of China (NSFC 12433004, U2031201), the Eighteenth Regular Meeting Exchange Project of the Scientific Technological Cooperation Committee between the People's Republic of China and the Republic of Bulgaria (Series No. 1802). JHF also acknowledges the science research grants from the China Manned Space Project with NO. CMS-CSST-2024-A07. MFG is supported by the National Science Foundation of China (grant 12473019), the China Manned Space Project with No. CMSCSST-2021-A06, the National SKA Program of China (Grant No.\ 2022SKA0120102), and the Shanghai Pilot Program for Basic Research-Chinese Academy of Science, Shanghai Branch (JCYJ-SHFY-2021-013). ZZ is funded by the National Science Foundation of China (grant no.\ 12233005). 

\newpage
\appendix
\section{Testing the effect of the spectral window and seasonal gaps}\label{gap}

Gaps in a light curve, particularly recurring seasonal gaps caused by observing constraints, maintenance, or weather, can strongly affect periodogram analyses by introducing aliasing features. A standard diagnostic for assessing such effects is the spectral window, constructed by replacing the observed fluxes or magnitudes with unity while retaining the original observation times. The spectral window for our data (\autoref{fig:window}) exhibits a peak near 365 days, as expected from annual observing gaps. Because this feature lies at one year, any candidate periodicity on comparable timescales, such as the $\sim$ 433-day signal, requires careful scrutiny to distinguish genuine variability from sampling artifacts.

To investigate this issue, we performed forward simulations designed to reproduce both the intrinsic variability and the sampling properties of the data. We will start with simulations for a sample case and then generalize the results to a broader category (see \autoref{fig_A1}). We first generated a synthetic light curve ($LC_1$) with 1-day sampling using the EMP13 method, assuming a red-noise power spectral density with spectral index $\alpha = 1.50$. A QPO with a characteristic period of $\sim$ 433-days was then added in the time domain as a weakly modulated sinusoidal component, leaving the underlying red-noise PSD unchanged outside the narrow frequency range of the QPO. The simulated light curve was subsequently resampled at the original observation times to produce $LC_2$, thereby imposing the same seasonal gaps and spectral window as the real data. The resulting $LC_2$ and its spectral window are shown in the top-left and top-right panels of \autoref{fig_A2}, respectively, with the latter matching that of the observed light curve.

\restartappendixnumbering
\setcounter{figure}{0}
\setcounter{table}{0}

\begin{figure}
    \centering
    \includegraphics[width=1.12\linewidth]{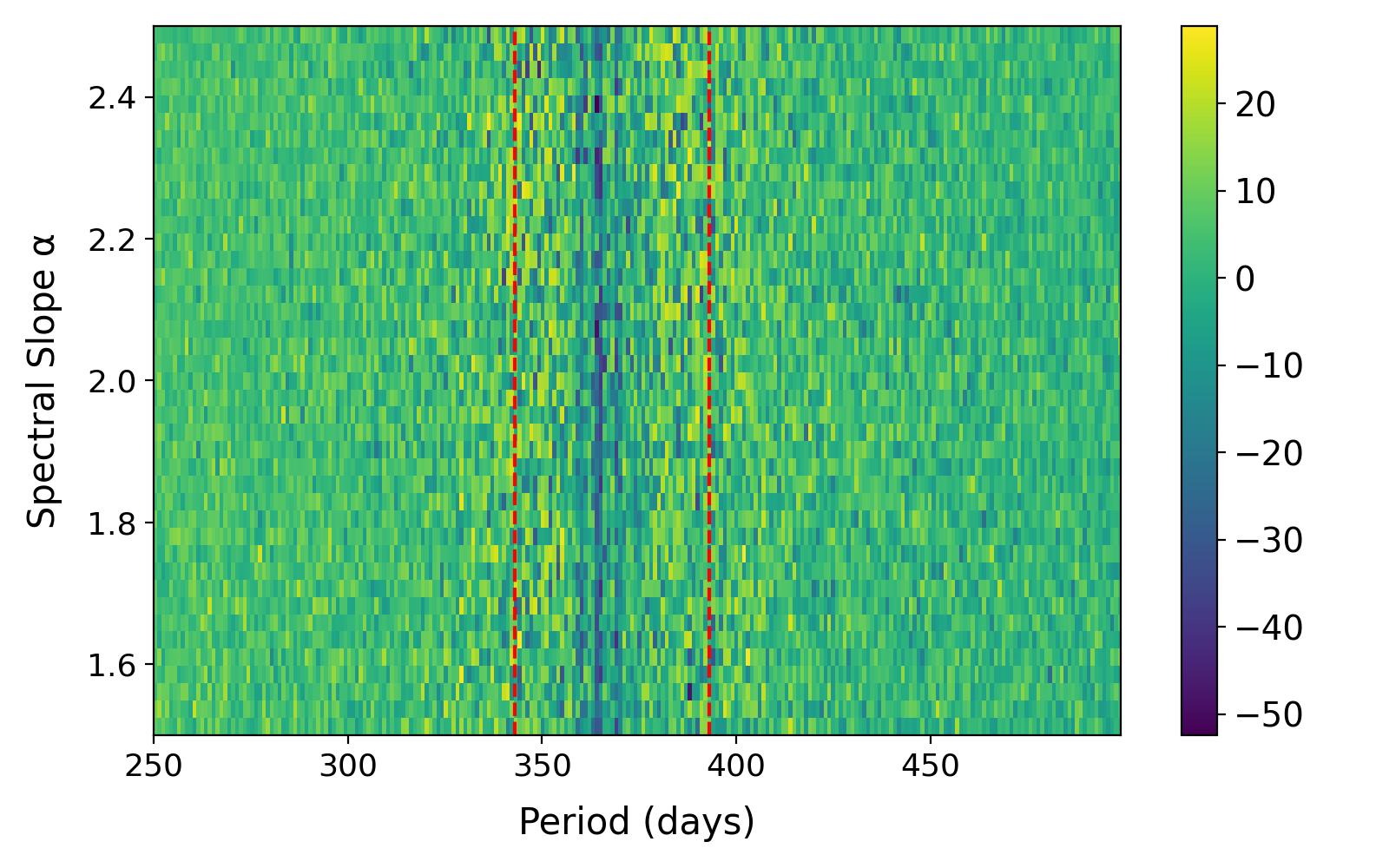}
    \caption{The figure shows the plot of the percentage power change after interpolation, with respect to the QPO periods and the spectral slopes of the simulated light curves. The red-dashed line represents a 1$\sigma$ exclusion region.}
    \label{fig_A1}
\end{figure}

\begin{figure*}
    \centering
    \includegraphics[width=0.95\linewidth]{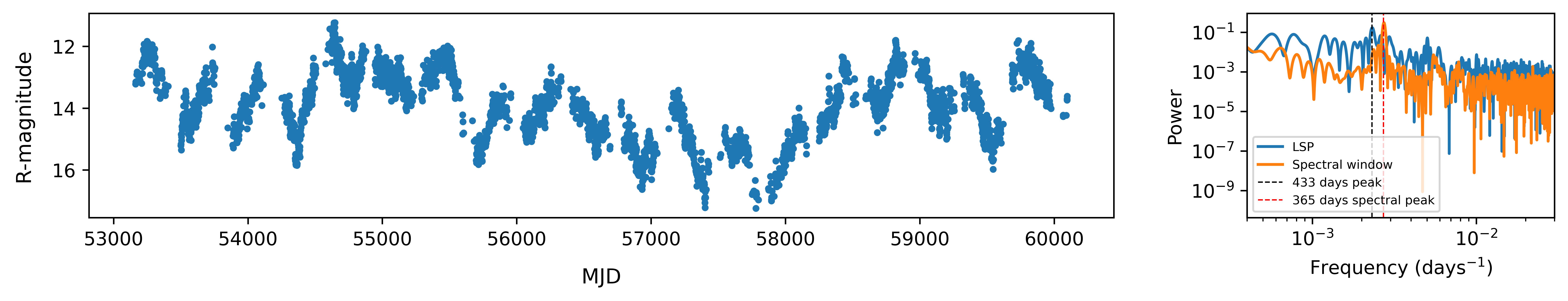}
    \includegraphics[width=0.95\linewidth]{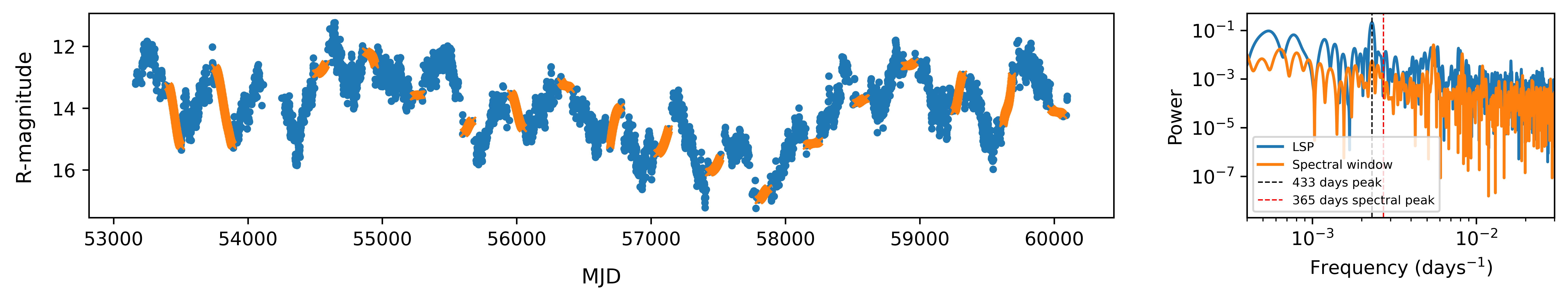}
    \caption{Top left: This figure displays an example of one of the simulated LCs with same timestamps as the original LC for $\alpha = 1.50$ having a QPO of 433 days. Top right: This plot shows the LSP of the simulated light curve (in blue) and its spectral window response (in orange). Bottom left: Similar to top left but with interpolated gaps shown in orange. Bottom right: Similar to top right, but for the light curve in bottom left, showing improvement in peak signal after interpolation.}
    \label{fig_A2}
\end{figure*}

To test the robustness of the detected periodicity, we applied an additional diagnostic based on gap interpolation. Specifically, we interpolated $LC_2$ across the seasonal gaps using a piecewise cubic Hermite interpolating polynomial (PCHIP). This interpolation scheme was chosen because it preserves local monotonicity and avoids the overshooting commonly introduced by higher-order spline methods. Importantly, interpolation is not used here to recover or enhance the signal, but rather as a diagnostic perturbation of the sampling window: if a periodogram peak is substantially influenced by the window function, modest modifications to the gaps should lead to a substantial change in its power. Conversely, a genuine QPO is expected to remain comparatively stable. $LC_2$ and interpolated $LC_2$, along with their respective periodograms, are shown in \autoref{fig_A2}. We find that, for both the observed light curve (\autoref{fig:window}) and the simulated $LC_2$ (\autoref{fig_A2}), the power of the $\sim$ 433-day peak changes by no more than $\sim10\%$ after interpolation, corresponding to a global significance change of $<5\%$ for the observed light curve.\\

To assess the generality of our results, we repeated the simulations across a range of spectral indices ($\alpha = 1.50$ to $2.50$) and injected QPO periods between 250 and 500 days. The resulting percentage change in power after interpolation is shown in \autoref{fig_A1}. We find that the power variation can reach around $50\%$, but these large changes are concentrated around 368 days. This effect occurs because the peak of the sampling window function coincides with these periods. To quantify this effect more conservatively, we fitted a Gaussian function to the median power-change values (over each spectral index). From this fit, we derived a $1\sigma$ period range of $368\pm23$ days, indicated by the red dashed lines in \autoref{fig_A1}. Periods falling within this interval should therefore be treated carefully when claiming a QPO detection in light curves with yearly seasonal gaps. Based on this analysis, we excluded the $1\sigma$ region from our calculations and found that $>98\%$ periods have a power change of less than $15\%$ for all spectral indices. Importantly, our detected QPO lies outside this affected range, placing it in a comparatively safe region of the period space. Finally, we estimated the corresponding change in global significance for the maximum-power variation across all simulations, which amounts to $\sim 17\%$. While this is not the universal threshold for the acceptable change in significance after interpolation, these results suggest that genuine QPOs are less sensitive to moderate perturbations of the sampling window, whereas window-induced features are expected to vary significantly more \citep{2025MNRAS.tmp.1751P}. The relatively small changes in the power and significance of the 433-day peak, observed both in the real data and in simulated light curves containing an injected QPO, therefore suggest that this feature is unlikely to be a by-product of the annual observing gaps. This conclusion is further supported by a recent work \citep{2025MNRAS.tmp.1751P}, which demonstrates that even extreme data loss with up to $70\%$ missing observations does not suppress the detection significance of a periodic signal in LSP analyses. Taken together, the combined use of the spectral window, targeted simulations, and gap-interpolation diagnostics provides a robust framework for distinguishing true year-scale QPOs, such as the $\sim$ 433-day signal identified in the present study, from sampling artifacts.

\newpage

\bibliography{Biblio}{}
\bibliographystyle{aasjournal}

\end{document}